\documentclass[12pt]{article}
\usepackage[a4paper, left=2cm,right=2cm]{geometry}
\usepackage{hyperref}
\usepackage{amsfonts} 
\usepackage{amsmath}
\usepackage{amssymb}
\usepackage{setspace}
\usepackage{color}

\usepackage[utf8,applemac]{inputenc}
\usepackage{tensor}
\usepackage{cite}
\usepackage{tikz}
\usepackage{graphicx,subfigure}
\usepackage[font=small, font+=it, format=hang, margin={1cm, 1cm}]{caption}
\graphicspath{{figure/}}

\usepackage{dcolumn}
\usepackage{bm}

\numberwithin{equation}{section}

\newcommand{\bea}{\begin{eqnarray}}
\newcommand{\eea}{\end{eqnarray}}
\newcommand{\be}{\begin{equation}}
\newcommand{\ee}{\end{equation}}
\def\nn{\nonumber}

\newcommand{\cT}{\mathcal{T}}
\newcommand{\cB}{\mathcal{B}}

\newcommand{\cW}{\mathcal{W}}

\newcommand{\cD}{\mathcal{D}}

\newcommand{\cH}{\mathcal{H}}

\newcommand{\cE}{\mathcal{E}}

\newcommand{\cF}{\mathcal{F}}

\renewcommand{\d}{\textrm{d}}

\begin{document}


\setcounter{tocdepth}{2}

\begin{titlepage}

\begin{flushright}\vspace{-3cm}
{\small
To be published in Chinese Physics C\\
May 18, 2023}
\end{flushright}
\vspace{0.5cm}

\begin{center}

{{ \LARGE{\bf{The Analytical Solutions of Equatorial Geodesic Motion in Kerr Spacetime\\ }}}}
\vspace{5mm}

\bigskip

\centerline{\large{\bf{Yan Liu$^{\dagger}$\footnote{email: yanliu@mail.bnu.edu.cn}}, {Bing Sun$^{\ddagger,*}$ \footnote{email: bingsun@bua.edu.cn\\ Corresponding author: Yan Liu and Bing Sun}}}}

\vspace{2mm}
\normalsize
\bigskip\medskip
\textit{{}$^\dagger$ Department of Physics, Yantai University, Yantai 264005, China} \\
\textit{{}$^\ddagger$ CAS Key Laboratory of Theoretical Physics, Institute of Theoretical Physics, Chinese Academy of Sciences, Beijing 100190, China} \\
\textit{{}$^*$ Department of Basic Courses, Beijing University of Agriculture, Beijing 102206, China}

\vspace{15mm}

\begin{abstract}
\noindent

{ The study of Kerr geodesics has a long history, particularly for those occurring within the equatorial plane, which is generally well-understood. However, upon comparison with the classification introduced by one of us \href{https://journals.aps.org/prd/abstract/10.1103/PhysRevD.105.024075}{[Phys. Rev. D 105, 024075 (2022)]}, it becomes apparent that certain classes of geodesics, such as trapped orbits, are still lacking analytical solutions. Thus, in this study, we provide explicit analytical solutions for equatorial timelike geodesics in Kerr spacetime, including solutions of trapped orbits, which capture the characteristics of special geodesics, such as the positions and conserved quantities of circular orbits, bound orbits, and deflecting orbits. Specifically, we determine the precise location at which retrograde orbits undergo a transition from counter-rotating to prograde motion due to the strong gravitational effects near the rotating black hole. Interestingly, we observe that for orbits with negative energy, the trajectory remains prograde despite the negative angular momentum.
Furthermore, we investigate the intriguing phenomenon of deflecting orbits exhibiting an increased number of revolutions around the black hole as the turning point approaches the turning point of the trapped orbit. Additionally, we find that only prograde marginal deflecting geodesics are capable of traversing through the ergoregion.
In summary, our findings present explicit solutions for equatorial timelike geodesics and offer insights into the dynamics of particle motion in the vicinity of a rotating black hole.
}

\end{abstract}


\end{center}

\end{titlepage}

\tableofcontents

\section{Introduction}

In recent years, the detection of gravitational waves from binary black hole mergers~\cite{LIGOScientific:2016aoc} and the upcoming gravitational wave detection missions such as LISA~\cite{LISA:2017pwj}, Taiji~\cite{Ruan:2018tsw}, and Tianqin~\cite{TianQin:2020hid} projects have underscored the urgency and significance of modeling the two-body problem in general relativity. Especially, the motion of a small body in extreme mass ratio inspirals can be treated as a perturbation of timelike geodesic motion, which can be effectively modeled using the self-force approach~\cite{Pound:2021qin}.
Accretion disks around rotating black holes are intimately connected to the innermost stable circular orbit (ISCO) or unstable circular geodesics~\cite{Page:1974he}. The investigation of plunging and deflecting timelike geodesics offers valuable insights into the Penrose process~\cite{Penrose:1971uk} and the two-body scattering problem ~\cite{Damour:2017zjx}. Furthermore, the recent interest in analyzing black hole images~\cite{Lee:2022rtg,Su:2022roo,Yan:2021ygy,Zhang:2023cuw,Kraniotis:2019ked} is closely tied to null and timelike Kerr geodesics.

The study of Kerr geodesics started since the Kerr solution has been found in 1963~\cite{Kerr:1963ud}. In 1968, Carter found an extra conserved quantity called the Carter constant~\cite{Carter:1968rr} during the geodesic motion in Kerr spacetime, which was further discussed in~\cite{Compere:2023alp}. The bound geodesics in Kerr spacetime were studied by Wilkins~\cite{Wilkins:1972rs}. In~\cite{Bardeen:1973tla}, Bardeen studied the timelike and null geodesics, which were further analyzed in~\cite{Vazquez:2003zm,Hackmann:2015ewa,Lammerzahl:2015qps}. The achievement on the Kerr geodesics in the early stage were reviewed by Chandrasekhar~\cite{1983grg1.conf....6C}.
Twenty years ago, Mino introduced Mino time, decoupled the radial and polar motion~\cite{Mino:2003yg}. The bound geodesics were revisited, and the explicit form of the geodesics in terms of fundamental orbital frequencies was given in~\cite{Schmidt:2002qk}. Later, these geodesics were further analyzed using separatrix and elliptic functions~\cite{Levin:2008yp,Fujita:2009bp,vandeMeent:2019cam}. It was proved that only trapped orbits are allowed for negative energy geodesic motion in the ergoregion~\cite{Vertogradov:2015hza}. In recent years, the geodesic motion in the near-horizon region of high-spin black holes has been analyzed~\cite{Hadar:2015xpa,Kapec:2019hro,Compere:2017hsi,Compere:2020eat}. Gralla and Lupsasca provided analytical solutions for null geodesics in Kerr spacetime~\cite{Gralla:2019ceu}. The full classification of timelike radial geodesic motion has been done in~\cite{Compere:2021bkk}. A new method for calculating inspirals from the innermost stable circular orbit (ISCO) was introduced in~\cite{Mummery:2022ana}. Analytical solutions for geodesics related to circular and innermost stable spherical orbits in the phase space have been obtained~\cite{Mummery:2023hlo,Dyson:2023fws}. Nevertheless, compared with the classification of the equatorial geodesics in ~\cite{Compere:2021bkk}, some classes of orbits, like trapped orbits associated with parameters governing bound and deflecting motion, still lack explicit analytical solutions.

In this paper, we revisit the equatorial timelike geodesic motion in Kerr spacetime. Specifically, we focus on the generic orbits related to the ``special'' orbits, such as circular, bound, and deflecting orbits, i.e., the geodesic classes in the region $\ell^u\leq\ell\leq\ell^s$ of the phase space in Figure 8 of~\cite{Compere:2021bkk}. The motion on the equatorial plane is mainly dominated by the radial motion, which is constrained by the radial potential. For the radial potential on the equatorial plane, there is always a root located at $r=0$. Setting the mass of the black hole and the particle $M=\mu=1$, the radial potential can be reduced to a cubic polynomial of $r$,
\bea
R(r)=(E^2-1)r^3+2r^2+(a^2(E^2-1)-\ell^2)r+2(\ell-aE)^2,\label{rpotential}
\eea
where $E$ and $\ell$ are the conserved energy and angular momentum, and $a$ denotes the spin of the black hole. For the orbits that plunge into the black hole, the angular momentum must satisfy~\cite{Compere:2021bkk}
\bea
\ell\le\ell_+=\frac{2Er_+}{a},
\eea
where $\ell_+$ is the angular momentum of the root structure that has one root touching the horizon.

Analyzing the roots of the radial potential and its derivatives, one can obtain the classification of the radial motion in the parameter space, which has been done in Figure 8 of~\cite{Compere:2021bkk}. For the convenience of the discussion, we introduce the notation of the root structures in Table \ref{table:notations} following~\cite{Compere:2021bkk}. 

\begin{table}[!tbh]    \centering
\begin{tabular}{|c|c|c|c|c|}\cline{1-2}\cline{4-5}
\rule{0pt}{13pt}\textbf{Notation} & \textbf{Denotes} & & \textbf{Notation} & \textbf{Denotes}\\\cline{1-2}\cline{4-5}
$\vert$ &  outer horizon & & $\bullet$ & simple roots (turning points) \\\cline{1-2}\cline{4-5}
$+$ & allowed region & &  $\bullet \hspace{-2pt}\bullet$ & double roots (circular orbits)\\\cline{1-2}\cline{4-5}
$-$ & disallowed region & & $\bullet\hspace{-4pt}\bullet\hspace{-4pt}\bullet$ & triple roots (ISCO)\\\cline{1-2}\cline{4-5}
$\rangle$ & radial infinity & & ${\vert \hspace{-5pt} \bullet }$ & roots touching the horizon\\\cline{1-2}\cline{4-5}
\end{tabular}\caption{Notations for the root structures on the equatorial plane.}\label{table:notations}
\end{table}

This paper is organized as follows. In Section \ref{sec3}, we discuss the geodesics related to the circular orbits. In Section \ref{sec4}, we present the geodesics associated with the bound and deflecting orbits. The marginal orbits are discussed in Section \ref{sec5}. Finally, we provide a summary of our results in Section \ref{sec6}. In Appendix \ref{appA}, we give the definition of elliptic integrals used in this paper. We prove that the solution of trapped orbits associated with bound and deflecting orbits can return to the stable and unstable cases in Appendix \ref{appB}. During the preparation of this paper, Adam, Eva and Patryk solved the non-equatorial Kerr geodesic motion in terms of Weierstrass elliptic functions \cite{Cieslik:2023qdc}. We compare the deflecting orbits with the non-equatorial results in \cite{Cieslik:2023qdc} as a consistency check in Appendix \ref{appC}.

\section{Orbits Related to Circular Orbits}\label{sec3}

In this section we suppose the circular orbit locates at $r_*$. The angular momentum and the energy of the circular orbits are obtained by solving for $R(r_*)=0$ and $R'(r_*)=0$,
\bea
\ell_{a,b}(E,r_*)&=&\frac{-2aE\mp\sqrt{r_*(2+(E^2-1)r_*)\Delta(r_*)}}{r_*-2},\\
E_{\pm}^{(1)}(r_*)&=&\pm\frac{(r_*-2)\sqrt{r_*}- a}{r_*^{3/4}\sqrt{(r_*-3)\sqrt{r}-2a}},\\
E_{\pm}^{(2)}(r_*)&=&\pm\frac{(r_*-2)\sqrt{r_*}+ a}{r_*^{3/4}\sqrt{(r_*-3)\sqrt{r}+2a}}.
\eea
The branches of the solution are:
\bea
(\ell_a(E_+^{(1)}),E_+^{(1)}), (\ell_a(E_-^{(2)}),E_-^{(2)}), (\ell_b(E_-^{(1)}),E_-^{(1)}), (\ell_b(E_+^{(2)}), E_+^{(2)});
\eea
Based on the analysis of ~\cite{Compere:2021bkk}, the allowed circular motion are $(\ell_a(E_+^{(1)}),E_+^{(1)})$ and $(\ell_b(E_+^{(2)}), E_+^{(2)})$, which can be reduced to 
\bea
\ell^{(1),(2)}(r_*)&=&\pm\frac{r_*^2\mp2 a \sqrt{r_*}+ a^2}{r_*^{3/4} \sqrt{r_*^{3/2}-3 \sqrt{r_*}\pm2 a}},\\
E^{(1),(2)}(r_*)&=&\frac{(r_*-2)\sqrt{r_*}\pm a}{r_*^{3/4}\sqrt{(r_*-3)\sqrt{r_*}\pm2a}},
\eea
where branch $(1)$ with upper sign denotes the prograde circular obits, and branch $(2)$ with lower sign denotes the retrograde orbits. 

Here we introduce the special values in~\cite{Compere:2017hsi}:
\bea
r_c^{(1)}&=&2-a+2\sqrt{1-a},\\
r_c^{(2)}&=&2+a+2\sqrt{1+a},\\
r_*^{(1)}&=&2+\cos(\frac{2\arcsin a}{3})-\sqrt{3}\sin(\frac{2\arcsin a}{3}),\\
r_*^{(2)}&=&2+\cos(\frac{2\arcsin a}{3})+\sqrt{3}\sin(\frac{2\arcsin a}{3})
\eea
where $r_c^{(1),(2)}$ are obtained by solving $E^{(1),(2)}=1$, and $r_*^{(1),(2)}$ are obtained by the condition $E^{(1),(2)}$ are real, such that at $r_*^{(1),(2)}$, we have $E^{(1),(2)}\to \infty$.

\paragraph{Circular orbits with Energy $|E|\neq1$}
Ignoring the ISCO, there are three types of root structures related to the circular orbits, $\vert+\bullet\hspace{-4pt}\bullet+\bullet-\rangle$ and $\vert+\bullet-\bullet\hspace{-4pt}\bullet-\rangle$ for $|E|<1$, and $\vert+\bullet\hspace{-4pt}\bullet+\rangle$ for $E>1$. The allowed orbits in the $+$ region have the same angular momentum and energy with the corresponding circular orbits. Note that there are no permitted orbits associated to circular orbits with energy $E\leq-1$.  Now consider the radial potential in a form as 
\bea
\frac{R(r)}{E^2-1}=(r-r_1)(r-r_*)^2,
\eea
comparing with \eqref{rpotential} and replacing $(E,\ell)$ by the angular momentum and energy of the prograde and retrograde circular orbits $(E^{(1),(2)},\ell^{(1),(2)})$, the another root can be obtained in terms of the position of the circular orbits $r_*$,
\bea
r_1=r_{1^\pm}=\frac{2 r_* \left(a\mp\sqrt{r_*}\right)^2}{-a^2\pm4 a \sqrt{r_*}+(r_*-4) r_*},
\eea
where the $+$ and $-$ indices denote the prograde and retrograde orbits. 

With the energy $|E|<1$, when $r_{1^\pm}=r_*=r_{I^\pm}$, the double root and the single root merge into a triple root, and turns out to be the innermost stable circular orbits (ISCO) $\vert+\bullet\hspace{-4pt}\bullet\hspace{-4pt}\bullet\hspace{2pt}-\rangle$ with the angular momentum and energy $(\ell_{I^{\pm}}, E_{I^{\pm}})$, which has been revisited in \cite{Mummery:2022ana}. When $r_+<r_{1}<r_*$, the circular orbits are stable with the root structure $\vert+\bullet-\bullet\hspace{-4pt}\bullet-\rangle$, the orbits in the $+$ region $r_+<r<r_{1}$ are trapped orbits $\cT^s$; when $r_{1}>r_*$, the circular orbits are unstable with the root structure $\vert+\bullet\hspace{-4pt}\bullet+\bullet-\rangle$, the orbits in the first $+$ region $r_+<r<r_*$ are whirling trapped orbits $\cW\cT^u$, and the orbits in the second $+$ region $r_*<r<r_{1}$ are whirling bound orbits $\cW\cB^u$, or homoclinic orbits $\cH^u$. 

With the energy $E>1$,  the another root $r_1<0$, but one can easily prove that we always have $-r_1>r_*$ for $r_*>r_+$. The circular orbits are unstable with the root structure $\vert+\bullet\hspace{-4pt}\bullet+\rangle$, the orbits in the first $+$ region $r_+<r<r_*$ are whirling trapped orbits $\cW\cT^u$, and the orbits in the second $+$ region $r>r_*$ are whirling deflecting orbits $\cW\cD^u$.

The $r$ and $\phi$ components of the 4-velocity of the orbits in the $+$ region can be expressed in terms of the parameters of related circular orbits,
\bea
U^r&=&\frac{\d r}{\d \tau}=\pm\frac{\sqrt{(E^2-1)r(r-r_{1})(r-r_*)^2}}{r^2},\\
U^{\phi}&=&\frac{\d\phi}{\d \tau}=\frac{2aE+\ell(r-2)}{r\Delta(r)},
\eea
where $\Delta(r)=(r-r_-)(r-r_+)$, the sign $+$ means the orbits are outgoing, and $-$ denotes the ingoing orbits. Particularly for the $\phi$ motion related to the radial motion, we have
\bea
\frac{\d\phi}{\d r}=\frac{\d\phi}{\d\tau}\frac{\d\tau}{\d r}=U^\phi\frac{\d\tau}{\d r}.\label{phir}
\eea
From now on, we only discuss the ingoing orbits, for the outgoing orbits, one can simply flip the sign by symmetry. 

\subsection{Unstable Circular Orbits}
\subsubsection{Whirling trapped and homoclinic orbits with $0<E<1$}
We first discuss the ingoing prograde orbits with the root structure $\vert+\bullet\hspace{-4pt}\bullet+\bullet-\rangle$. Then the energy is confined in the region $E_{I^\pm}<E^{(1),(2)}<1$, and the location of the unstable circular orbits are confined in the region $r_c^{(1),(2)}<r_*<r_{I^\pm}$.

For the whirling trapped orbits  ($\cW\cT^u$) related to the unstable circular orbits, which are the allowed motion in the first $+$ region $r_+<r<r_*<r_{1}$ asymptotically approach the unstable circular orbits. For $\cW\cT^u$ and the orbits inside the horizon $0<r<r_+$, the ingoing radial velocity is given by
\bea
U^r=\frac{\d r}{\d\tau}=-\frac{(r_*-r)\sqrt{(E^2-1)r(r-r_{1})}}{r^2},\label{WTuEm1}
\eea
which can be reorganized as
\bea
-\sqrt{1-E^2}\d\tau=\frac{1}{(\frac{r_*}{r}-1)\sqrt{\frac{r_{1}}{r}-1}}\d r.
\eea

The homoclinic orbits ($\cH^u$) are related to the unstable circular orbits, which is the allowed motion in the second $+$ region $r_*<r<r_{1}$, and asymptotically approach the unstable circular orbits. The radial velocity of $\cH^u$ is

\bea
U^r=-\frac{(r-r_*)\sqrt{(E^2-1)r(r-r_{1})}}{r^2}.\label{HuEm1}
\eea

After integrate \eqref{HuEm1} in the corresponding region, we obtain the solution of the proper time of $\cH^u$ in the region $r_*<r<r_1$,
\bea
-\sqrt{1-E^2}\tau=r\sqrt{\frac{r_1}{r}-1}+(r_1+2r_*)\arctan\sqrt{\frac{r_1}{r}-1}+\frac{2r_*^{3/2}}{\sqrt{r_1-r_*}}\tanh^{-1}\sqrt{\frac{r(r_1-r_*)}{r_*(r_1-r)}}.
\eea
Note that it is not real in the region $r<r_*$ due to the $\tanh^{-1}$ function. However, by utilizing the property of this function: 
\bea
\frac{\d}{\d x}\tanh^{-1}(x)=\frac{\d}{\d x}\tanh^{-1}(1/x),
\eea
we can adjust the solution and obtain the proper time of the orbits in the region $r<r_*$,
\bea
\sqrt{1-E^2}\tau=r\sqrt{\frac{r_1}{r}-1}+(r_1+2r_*)\arctan\sqrt{\frac{r_1}{r}-1}+\frac{2r_*^{3/2}}{\sqrt{r_1-r_*}}\tanh^{-1}\sqrt{\frac{r_*(r_1-r)}{r(r_1-r_*)}}.
\eea
For the $\phi$ motion, by integrating \eqref{phir} in corresponding region and setting the integral constants $\phi_0=0$, we have the solutions of the orbits in the following
\begin{itemize}
    \item whirling trapped orbits in the region $r_+<r<r_*$,
        \bea
        \phi=C_*^1\tanh^{-1}\sqrt{\frac{r(r_1-r_*)}{r_*(r_1-r)}}+C_-^1\tanh^{-1}\sqrt{\frac{r_-(r_1-r)}{r(r_1-r_-)}}+C_+^1\tanh^{-1}\sqrt{\frac{r_+(r_1-r)}{r(r_1-r_+)}},\label{WTphi}
        \eea
    \item homoclinic orbits in the region $r_*<r<r_1$,
        \bea
        \phi=-C_*^1\tanh^{-1}\sqrt{\frac{r_*(r_1-r)}{r(r_1-r_*)}}-C_-^1\tanh^{-1}\sqrt{\frac{r_-(r_1-r)}{r(r_1-r_-)}}-C_+^1\tanh^{-1}\sqrt{\frac{r_+(r_1-r)}{r(r_1-r_+)}},
        \eea
    \item the orbits in the region $r_-<r<r_+$,
        \bea
        \phi=C_*^1\tanh^{-1}\sqrt{\frac{r(r_1-r_*)}{r_*(r_1-r)}}+C_-^1\tanh^{-1}\sqrt{\frac{r_-(r_1-r)}{r(r_1-r_-)}}+C_+^1\tanh^{-1}\sqrt{\frac{r(r_1-r_+)}{r_+(r_1-r)}},
        \eea
    \item the orbits in the region $0<r<r_-$,
        \bea
        \phi=C_*^1\tanh^{-1}\sqrt{\frac{r(r_1-r_*)}{r_*(r_1-r)}}+C_-^1\tanh^{-1}\sqrt{\frac{r(r_1-r_-)}{r_-(r_1-r)}}+C_+^1\tanh^{-1}\sqrt{\frac{r(r_1-r_+)}{r_+(r_1-r)}},
        \eea
\end{itemize}
where the constants read as
\bea
C_*^1=\frac{2\sqrt{r_*}(2(\ell-aE)-r_*\ell)}{\sqrt{1-E^2}\sqrt{r_1-r_*}(r_*-r_-)(r_*-r_+)},\\
C_-^1=\frac{2\sqrt{r_-}(2(\ell-aE)-r_-\ell)}{\sqrt{1-E^2}\sqrt{r_1-r_-}(r_*-r_-)(r_+ - r_-)},\\
C_-^1=\frac{2\sqrt{r_+}(2(\ell-aE)-r_+\ell)}{\sqrt{1-E^2}\sqrt{r_1-r_+}(r_*-r_+)(r_+ - r_-)}.
\eea

\begin{figure}
     \centering
     \includegraphics[width=0.9\textwidth]{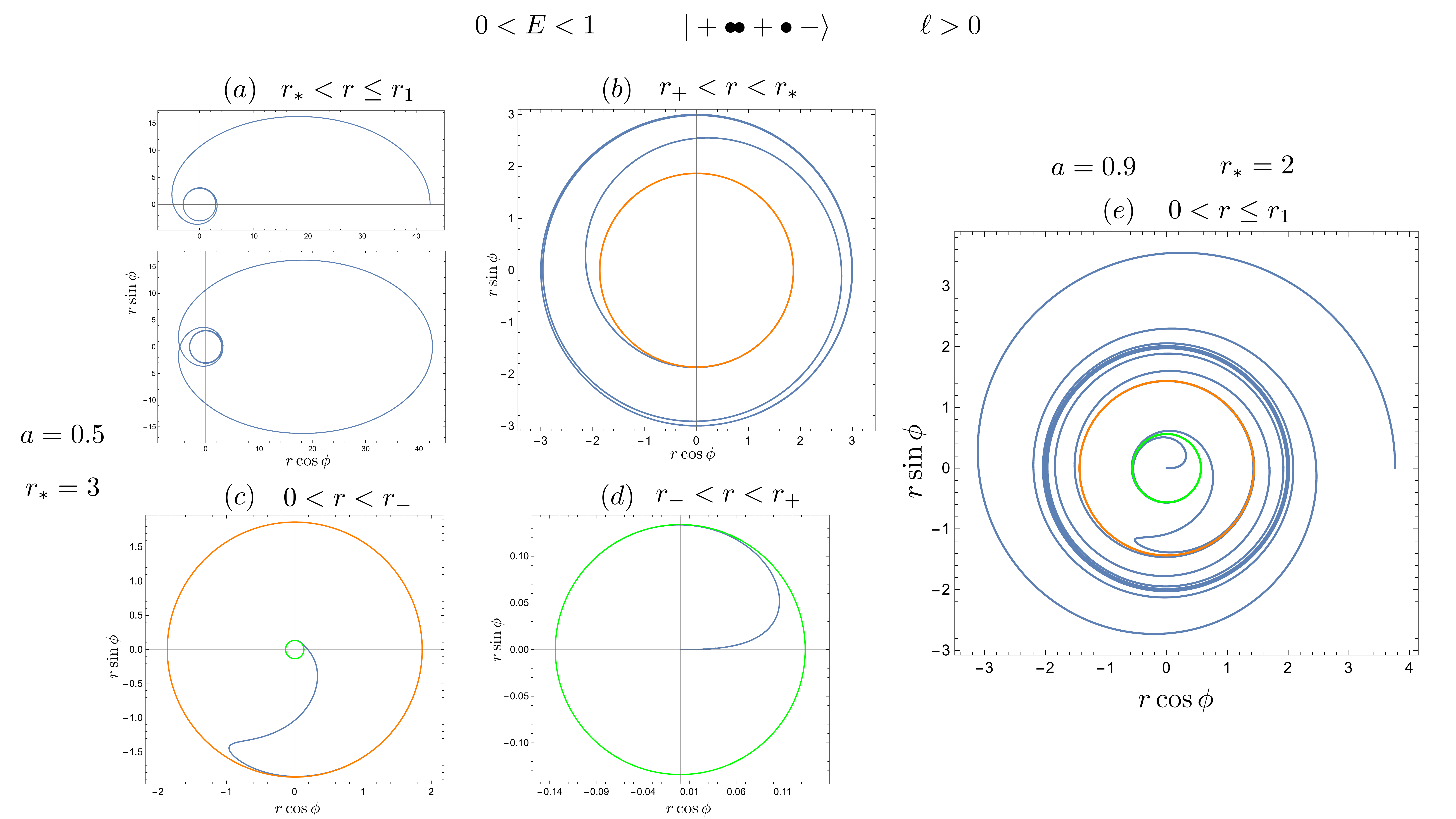}
     \caption{Geodesics associated with unstable prograde circular orbits. The blue lines illustrate the orbits, while the orange line denotes the outer horizon and the green line represents the inner horizon. The color scheme remains consistent throughout the subsequent figures.}
     \label{UnCirEm1}
 \end{figure}

In Figure \ref{UnCirEm1}, we illustrate the geodesics associated with prograde unstable circular orbits, where $0<E<1$. The behavior of different orbit classes in various regions is shown separately in plots $(a)$ to $(d)$. Plot $(a)$ demonstrates the behavior of ingoing and outgoing-to-ingoing homoclinic orbits confined within the region $r_*<r\leq r_1$, which asymptotically approach the unstable circular orbits. This is consistent with previous findings~\cite{Levin:2008yp}. Plot $(b)$ depicts a trapped orbit that originates from the unstable circular orbits and eventually plunges into the black hole. Plots $(c)$ and $(d)$ illustrate the trajectories from $r_+$ to $r_-$ and from $r_-$ to $0$, respectively. It is notable that the direction of motion changes at the horizons due to the presence of $\Delta(r)$ in the denominator of $U^\phi$, despite the conserved energy and angular momentum remaining constant. Finally, in plot $(e)$, we present the entire range of geodesics, choosing $a=0.9$ to ensure visibility inside the horizon.

\begin{figure}
     \centering
     \includegraphics[width=0.9\textwidth]{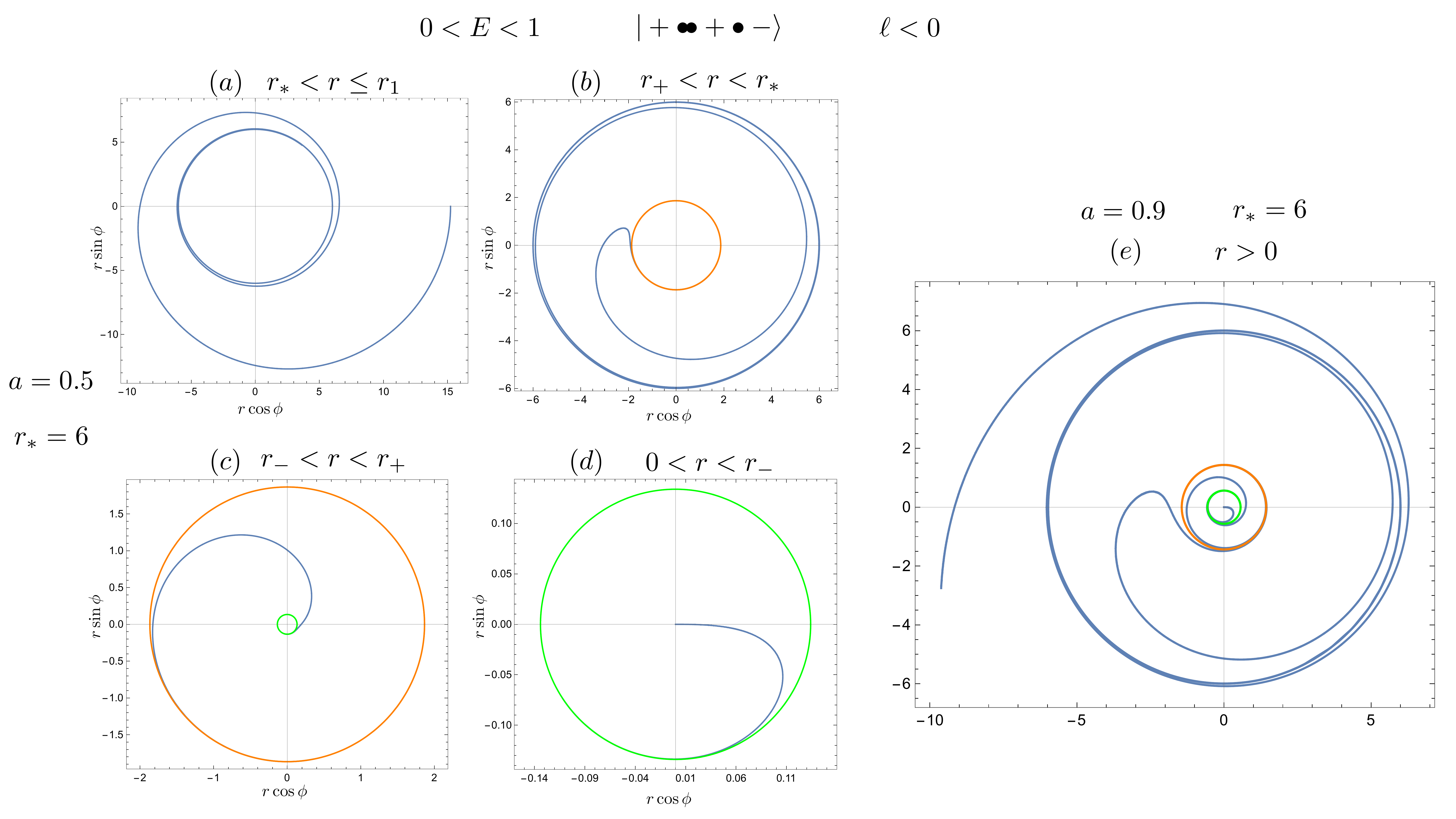}
     \caption{Geodesics related to the unstable retrograde circular orbits with the root structure $\vert+\bullet\hspace{-4pt}\bullet+\bullet-\rangle$.}
     \label{UnReCirEm1}
 \end{figure}

In Figure\ref{UnReCirEm1} we present the geodesics related to the retrograde unstable circular orbits with energy $0<E<1$ Plot $(b)$ shows an interesting phenomenon that near the horizon the strong gravity drags the retrograde trajectory to become prograde. As a result, the location of the turning point of the $\phi$ motion outside the horizon,
\bea
r_{\phi T}=2+\frac{2aE}{\ell}, \label{turningphi}
\eea
depends on the position of the circular orbits $r_*$ and the spin of the black hole $a$, which can be obtained by solving $\frac{\d\phi}{\d r}=0$.

It should be noted that \eqref{turningphi} is applicable not only to circular orbits but also to other retrograde trapped orbits, including those associated with bound or deflecting orbits or purely trapped orbits with the root structure $\vert+\bullet\hspace{2pt}-\rangle$. By inputting the corresponding energy and angular momentum values, one can obtain explicit expressions for these orbits as well.

\subsubsection{Whirling trapped and deflecting orbits with $E>1$}
When $E>1$, the circular orbits are unstable with the root structure $\vert+\bullet\hspace{-4pt}\bullet+\rangle$. The orbits in the first $+$ region are the whirling trapped orbits, and in the second $+$ region are whirling deflecting orbits.

For the $\cW\cT^u$ related to the unstable circular orbits in the region $r_+<r<r_*$ and the orbits inside the horizon $0<r<r_+$, the ingoing radial velocity is given by 
\bea
U^r=\frac{\d r}{\d\tau}=-\frac{(r_*-r)\sqrt{(E^2-1)r(r-r_1)}}{r^2}.
\eea
In the region $r>r_*$, the radial velocity for the $\cW\cD^u$ is given by
\bea
U^r=\frac{\d r}{\d\tau}=-\frac{(r-r_*)\sqrt{(E^2-1)r(r-r_1)}}{r^2}.
\eea
After the integration in the corresponding region, we obtain the proper time of the orbits in the region $r<r_*$
\bea
-\sqrt{1-E^2}\tau=\frac{2r_*^{3/2}}{\sqrt{r_*-r_1}}\tanh^{-1}\sqrt{\frac{r(r_*-r_1)}{r_*(r-r_1)}}-(r_1+2r_*)\sinh^{-1}\sqrt{\frac{r}{-r_1}}-\sqrt{r(r-r_1)},
\eea
and in the region $r>r_*$
\bea
\sqrt{1-E^2}\tau=\frac{2r_*^{3/2}}{\sqrt{r_*-r_1}}\tanh^{-1}\sqrt{\frac{r_*(r-r_1)}{r(r_*-r_1)}}-(r_1+2r_*)\sinh^{-1}\sqrt{\frac{r}{-r_1}}-\sqrt{r(r-r_1)}.
\eea

While the $\phi$ motion can be obtained by integrating \eqref{phir} in the corresponding region as follows:\\
\begin{itemize}
    \item whirling trapped orbits in the region $r_+<r<r_*$,
        \bea
        \phi=C_*^2\tanh^{-1}\sqrt{\frac{r(r_*-r_1)}{r_*(r-r_1)}}+C_-^2\tanh^{-1}\sqrt{\frac{r_-(r-r_1)}{r(r_--r_1)}}+C_+^2\tanh^{-1}\sqrt{\frac{r_+(r-r_1)}{r(r_+-r_1)}},\label{WTUnCir}
        \eea
    \item whirling deflecting orbits in the region $r>r_*$,
        \bea
        \phi=-C_*^2\tanh^{-1}\sqrt{\frac{r_*(r-r_1)}{r(r_*-r_1)}}-C_-^2\tanh^{-1}\sqrt{\frac{r_-(r-r_1)}{r(r_--r_1)}}-C_+^2\tanh^{-1}\sqrt{\frac{r_+(r-r_1)}{r(r_+-r_1)}},
        \eea
    \item orbits in the region $r_-<r<r_+$,
        \bea
        \phi=C_*^2\tanh^{-1}\sqrt{\frac{r(r_*-r_1)}{r_*(r-r_1)}}+C_-^2\tanh^{-1}\sqrt{\frac{r_-(r-r_1)}{r(r_--r_1)}}+C_+^2\tanh^{-1}\sqrt{\frac{r(r_+-r_1)}{r_+(r-r_1)}},
        \eea
    \item orbits in the region $0<r<r_-$,
        \bea
        \phi=C_*^2\tanh^{-1}\sqrt{\frac{r(r_*-r_1)}{r_*(r-r_1)}}+C_-^2\tanh^{-1}\sqrt{\frac{r(r_--r_1)}{r_-(r-r_1)}}+C_+^2\tanh^{-1}\sqrt{\frac{r(r_+-r_1)}{r_+(r-r_1)}},
        \eea
\end{itemize}
where the constants is given by
\bea
C_*^2=\frac{2\sqrt{r_*}(2(\ell-aE)-r_*\ell)}{\sqrt{1-E^2}\sqrt{r_*-r_1}(r_*-r_-)(r_*-r_+)},\\
C_-^2=\frac{2\sqrt{r_-}(2(\ell-aE)-r_-\ell)}{\sqrt{1-E^2}\sqrt{r_--r_1}(r_*-r_-)(r_+ - r_-)},\\
C_-^2=\frac{2\sqrt{r_+}(2(\ell-aE)-r_+\ell)}{\sqrt{1-E^2}\sqrt{r_+-r_1}(r_*-r_-)(r_- - r_+)}.
\eea

\begin{figure}
     \centering
     \includegraphics[width=0.9\textwidth]{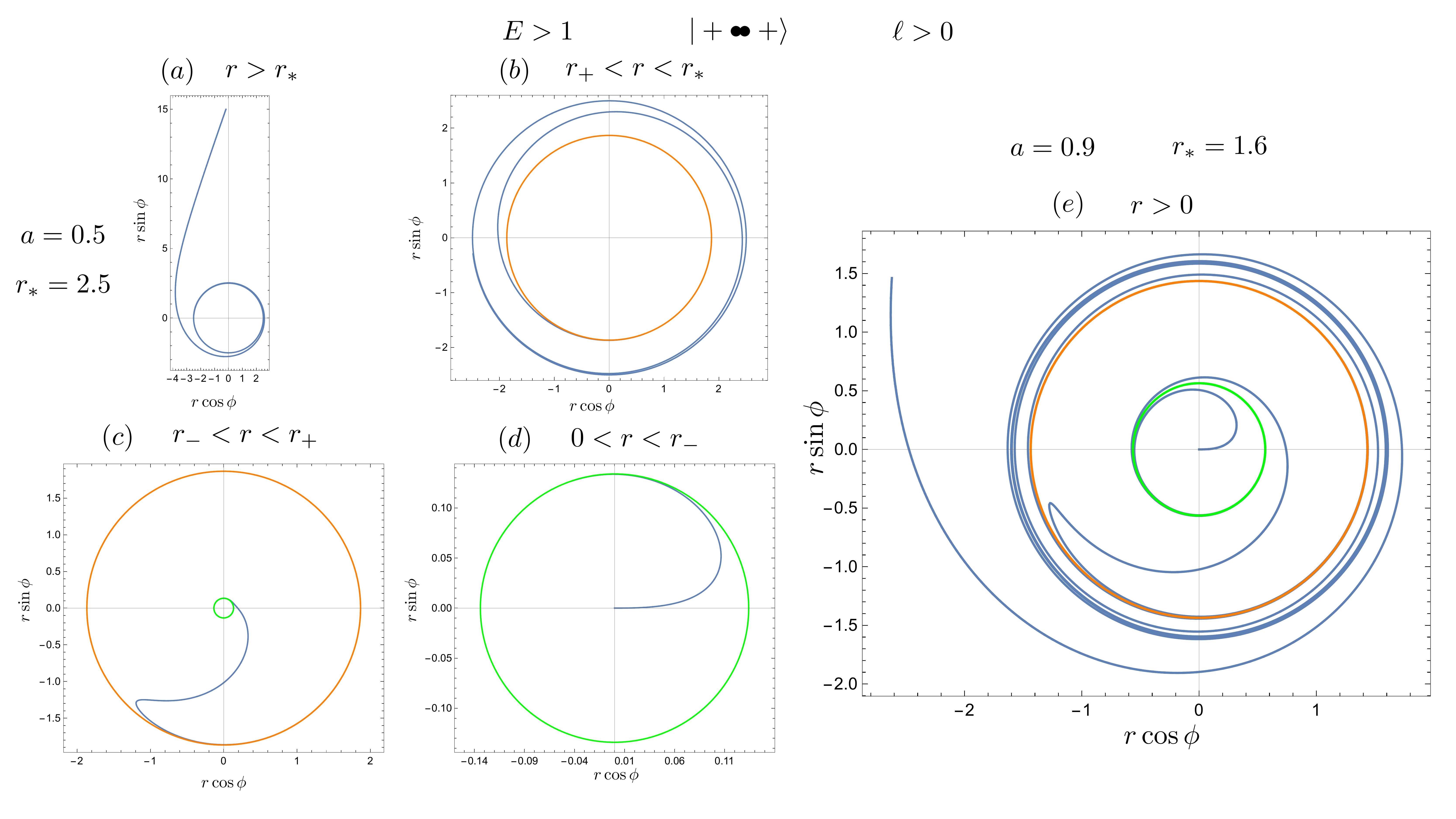}
     \caption{Geodesics related to the unstable prograde circular orbits with $E>1$.}
     \label{UnCirEp1}
 \end{figure}

\begin{figure}
     \centering
     \includegraphics[width=0.9\textwidth]{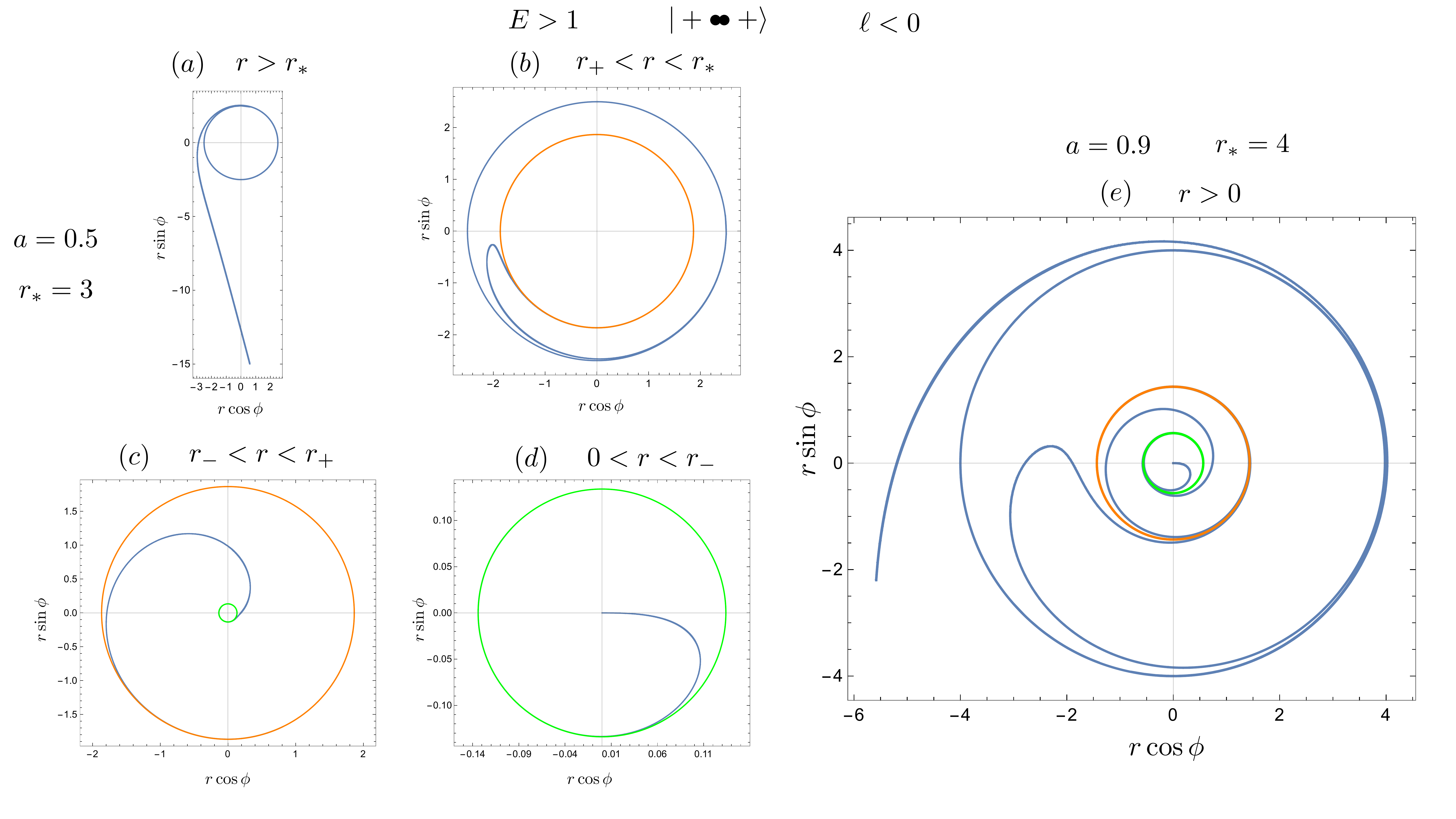}
     \caption{Geodesics related to the unstable retrograde circular orbits with $E>1$.}
     \label{UnReCirEp11}
 \end{figure}

In Figure~\ref{UnCirEp1} and Figure~\ref{UnReCirEp11}, we illustrate the geodesics associated with the prograde and retrograde unstable circular orbits with energy $E>1$. Plots $(a)$ to $(d)$ display the distinct classes of orbits, where $a=0.5$ and $r_*=3$. Additionally, plot $(e)$ showcases all classes of orbits within the root structure $\vert+\bullet\hspace{-4pt}\bullet\hspace{2pt}+\rangle$, with $a=0.9$ and $r_*=4$. Notably, plots $(a)$ demonstrate the whirling deflecting orbits originating from far infinity and asymptotically approaching the unstable circular orbits. The turning points of the trapped retrograde orbits, located at $r=r_{\phi T}$, are also clearly evident.

\subsection{Stable Circular Orbits}

We now discuss the ingoing trapped orbits with the root structure $\vert+\bullet-\bullet\hspace{-4pt}\bullet-\rangle$. These orbits are characterized by having energy values confined in the region  $E_{I^\pm}<E<1$ and stable circular orbits located at $r_*>r_{I^\pm}$. It is important to note that when $r_{1^+}=r_+$, the single root touches the horizon, and the stable circular orbit is located at 
\bea
r_*=r^s_+=\frac{4r_+}{a^2}+2\sqrt{3+r_+ +\frac{8r_+}{a^4}-\frac{4(2r_+ +1)}{a^2}}-r_+ -2.
\eea
At this point, the angular momentum and the energy are given by
\bea
\ell&=&\ell^{(1)}=\ell_+,\\
E&=&E^{(1)}=\frac{a\ell^{(1)}}{2r_+}.
\eea
When $r_*>r^s_+$, the angular momentum is greater than $\ell_+$, indicating that although the stable circular orbits exist, the trapped orbits are disallowed. However, for the orbital motion with negative energy, the trapped orbits exist even though the related circular orbits are disallowed. In this scenario, we can employ the parameters of the circular orbits to investigate the motion with negative energy, taking advantage of the symmetry in $R(r)$ achieved by flipping the sign of $E$ and $\ell$ simultaneously.

Based on the analysis above, we can draw the following conclusions regarding the existence of trapped orbits:
\begin{itemize}
\item For prograde orbits, the trapped orbits exist within the range $r_c^{(1)}<r_*<r_+^s$.
\item For retrograde orbits, the trapped orbits exist when $r_*>r_c^{(2)}$.
\item For orbits with negative energy, the trapped orbits exist when $r_*>r^s_+$.
\end{itemize}
These findings offer valuable insights into the critical role of $r_*$ in determining the presence of trapped orbits alongside the associated stable circular orbits. These insights are particularly useful when plotting the trajectories of these trapped orbits. To ensure the existence of trapped orbits, it is crucial to carefully select appropriate values for $r_*$. By doing so, one can accurately depict the trajectories and study the characteristics of these intriguing orbits.

For the trapped orbits related to the stable circular orbits, which represent allowed motion in the $+$ region $r_+<r<r_1<r_*$ of the root structure, as well as the orbits inside the horizon, the radial velocity is given by
\bea
U^r=-\frac{(r_*-r)\sqrt{(E^2-1)r(r-r_{1})}}{r^2}.
\eea
By integrating the above expression, we obtain the proper time as
\bea
-\sqrt{1-E^2}\tau=r\sqrt{\frac{r_1}{r}-1}+(r_1+r_*)\arctan\sqrt{\frac{r_1}{r}-1}-\frac{2r_*^{3/2}}{\sqrt{r_*-r_1}}\arctan\sqrt{\frac{r_*(r_1-r)}{r(r_*-r_1)}}.
\eea
For the azimuthal motion, we have the following expressions:\\
\begin{itemize}
    \item trapped orbits in the region $r_+<r\leq r_1$,
        \bea
        \phi=-C_*^2\arctan\sqrt{\frac{r_*(r_1-r)}{r(r_*-r_1)}}+C_-^1\tanh^{-1}\sqrt{\frac{r_-(r_1-r)}{r(r_1-r_-)}}+C_+^1\tanh^{-1}\sqrt{\frac{r_+(r_1-r)}{r(r_1-r_+)}}.\label{ScirTr}
        \eea
    \item the motion in the region $r_-<r<r_+$,
        \bea
        \phi=-C_*^2\arctan\sqrt{\frac{r_*(r_1-r)}{r(r_*-r_1)}}+C_-^1\tanh^{-1}\sqrt{\frac{r_-(r_1-r)}{r(r_1-r_-)}}+C_+^1\tanh^{-1}\sqrt{\frac{r(r_1-r_+)}{r_+(r_1-r)}}.
        \eea
    \item the motion in the region $0<r<r_-$,
        \bea
        \phi=-C_*^2\arctan\sqrt{\frac{r_*(r_1-r)}{r(r_*-r_1)}}+C_-^1\tanh^{-1}\sqrt{\frac{r(r_1-r_-)}{r_-(r_1-r)}}+C_+^1\tanh^{-1}\sqrt{\frac{r(r_1-r_+)}{r_+(r_1-r)}}.
        \eea
\end{itemize}

\begin{figure}
     \centering
     \includegraphics[width=0.99\textwidth]{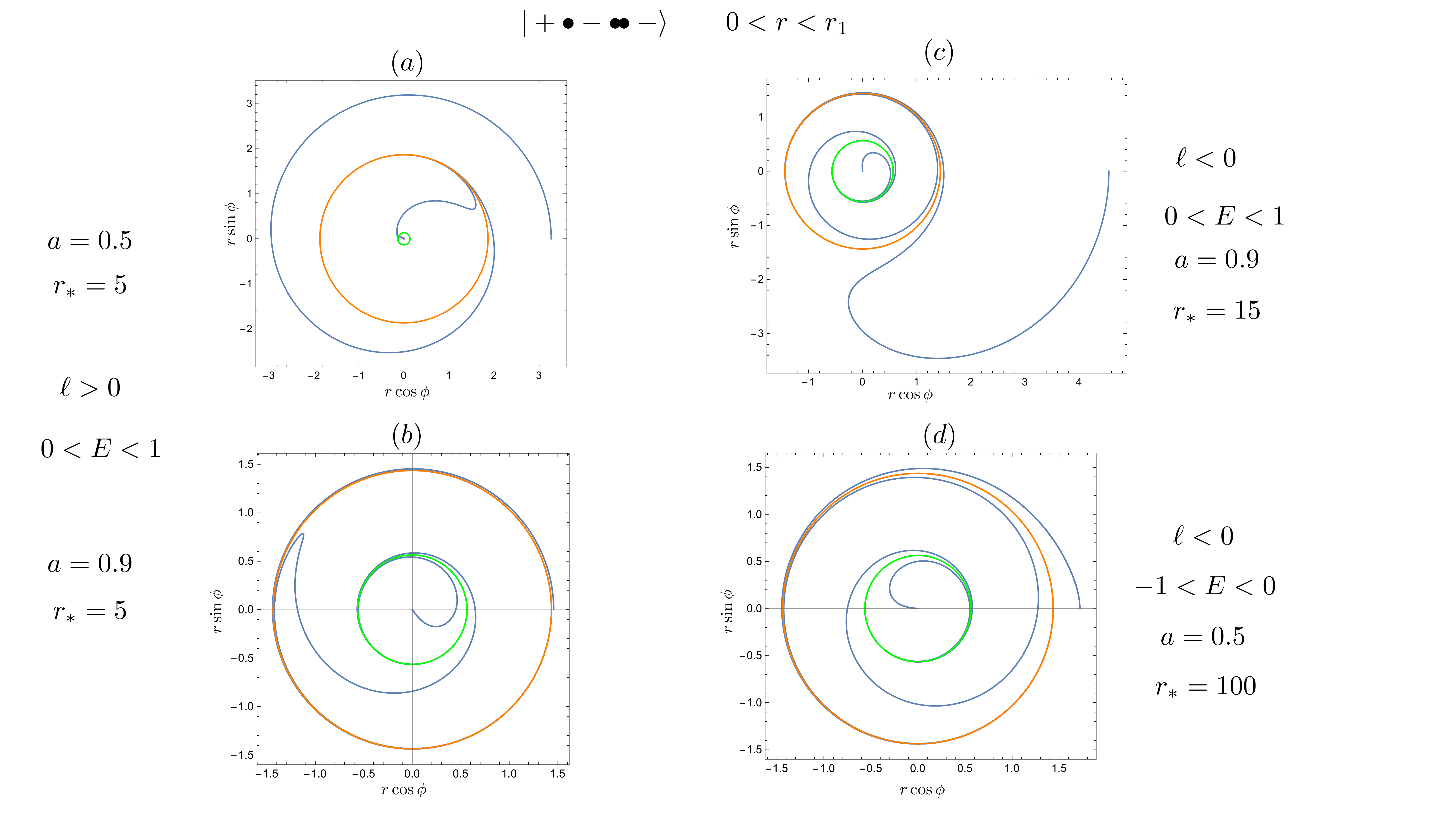}
     \caption{Geodesics related to the stable circular orbits with the root structure $\vert+\bullet-\bullet\hspace{-4pt}\bullet-\rangle$.}
     \label{SCir}
 \end{figure}

In Figure~\ref{SCir}, we present the geodesics in the region $0<r\leq r_1$ associated with stable circular orbits. Plots $(a)$ and $(b)$ display the geodesics related to prograde stable circular orbits with different black hole spins. It indicates that as the black hole rotates faster, the turning point $r_1$ approaches closer to the horizon. Plot $(c)$ illustrates the geodesics related to retrograde stable circular orbits. It is worth noting that for retrograde stable circular orbits, $r_*$ can go to infinity without affecting the existence of trapped orbits. Finally, plot $(d)$ showcases the geodesic motion with negative energy. Here, we observe that there are no turning points for the $\phi$ motion, both inside and outside the black hole. Furthermore, despite the angular momentum being negative, the trajectory remains prograde. One can check that our result match with \cite{Mummery:2023hlo} very well by replacing our notation in Eq.(53-54), Eq.(56-57), and Eq.(61-63) in \cite{Mummery:2023hlo}.

\section{Trapped Orbits with Separated Roots}\label{sec4}

In the region between $\ell^s$ and $\ell^u$ as shown in Figure 8 of~\cite{Compere:2021bkk}, the double root separate into two single roots. These single roots correspond to the turning points of bound orbits when $E<1$, and they represent the turning points of the trapped and deflecting orbits when $E>1$. The separation of these roots, which is described by the separatrix in terms of semilatus rectum and eccentricity, has been extensively discussed in previous works such as \cite{Glampedakis:2002ya,OShaughnessy:2002tbu,Levin:2008yp,vandeMeent:2019cam,Compere:2021bkk}. In this section, we delve into the analysis of trapped orbits associated with these separated roots and explore the properties of deflecting orbits. 

We first consider the root structure $\vert+\bullet-\bullet+\bullet\hspace{2pt}-\rangle$ with bound orbits. The trapped orbits in the first $+$ region $r_+<r<r_1$ are allowed when $\ell^u<\ell<\min(\ell^s,\ell^+)$ and $E_I<E<1$. When $\ell>\ell_+$, the trapped orbits are disallowed even though the bound orbits exist. Now we express the radial potential as 
\bea
\frac{R(r)}{E^2-1}=(r-r_1)(r-r_p)(r-r_a),\label{BD123}
\eea
where the roots of the radial potential $r_1<r_p<r_a$, representing the turning points of trapped and bound orbits. Here, $r_p=\frac{p}{1+e}$ and $r_a=\frac{p}{1-e}$ denote the pericenter and apocenter of the bound orbits respectively, where $p$ is the semilatus rectum and $e$ is the eccentricity. By comparing the coefficients of $r$ with \eqref{rpotential}, one can obtain the following solution,
\bea
r_1&=&\frac{2}{1-E^2}-(r_a+r_p)\label{r1bound}, \\
\ell^{(1),(2)}&=&aE\pm\sqrt{\frac{r_a r_p(2+(E^2-1)(r_a+r_p))}{2}},\label{Lbound}\\
E^{(1),(2)}&=&\sqrt{\frac{X_1+X_2\pm4\sqrt{2}a\sqrt{r_a r_p(r_a+r_p)\Delta(r_a)\Delta(r_p)}}{X_3+X_4}}.\label{Ebound}
\eea
where the indices $(1)$ and $(2)$ correspond to the $+$ and $-$ sign in $\pm$, and 
\bea
X_1&=&(r_a-2)(r_p-2)(r_a+r_p)(r_a(r_a+r_p)(r_p-2)-2r_p^2),\\
X_2&=&a^2(r_a^2(4-6r_p)-6r_a r_p(r_p-2)+4r_p^2),\\
X_3&=&r_a^3(r_a+2r_p )(r_p-2)^2+4r_p^2(r_p^2-r_a (\Delta(r_p)+a^2)),\\
X_4&=&r_a^2 r_p(r_p(r_p-6)(r_p-2)-8a^2).
\eea
Note that we have $\ell^{(1)}>0$, $\ell^{(2)}<0$. The allowed branches are as follows: the prograde orbits $(\ell^{(1)}(E^{(2)}),E^{(2)})$, the retrograde orbits $(\ell^{(2)}(E^{(1)}),E^{(1)})$, and the orbits with negative energy $(\ell^{(2)}(-E^{(2)}),-E^{(2)})$.

By replacing $r_p=\frac{p}{1+e}$ and $r_a=\frac{p}{1-e}$ in \eqref{BD123}, and compare the coefficients of $r$ with the radial potential, one can obtain the quantities and the polynomial that the semi-latus rectum and eccentricity should satisfy,
\bea
E&=&\pm\sqrt{1+\frac{2(e^2-1)}{2p+r_1(1-e^2)}}\\
\ell&=&aE\pm p\sqrt{\frac{r_1}{2p+(1-e^2)r_1}}\\
0&=&(p(4r_1-p(r_1-2))-a^2(2p+r_1(1-e^2)))^2\nn\\
&-&4a^2 p^2 r_1(r_1+2(p-1)-e^2(r_1-2))
\eea

In the case of trapped and deflecting orbits with the root structure $\vert+\bullet-\bullet\hspace{2pt}+\rangle$, $r_a$ is negative such that no longer be the apocenter, which we denote by $r_n$, and $r_p$ is no longer the pericenter, instead as the turning point of the deflecting orbits denoted by $r_d$. Note that we still have $r_n=\frac{p}{1-e}$ and $r_d=\frac{p}{1+e}$, and $e>1$ is no longer the eccentricity now. The another single root $r_1$, and the energy and angular momentum of the orbits can be obtained by simply replacing $r_a$ and $r_p$ with $r_n$ and $r_d$ in \eqref{r1bound}-\eqref{Ebound}.

\subsection{Trapped Orbits Related to Bound Orbits}
The radial velocity of the trapped orbit associated with a bound orbit is given by
\bea
U^r=-\frac{\sqrt{(E^2-1)r(r-r_1)(r-r_p)(r-r_a)}}{r^2},
\eea
which can be rewritten as
\bea
-\sqrt{(1-E^2)}\d\tau=\frac{1}{\sqrt{(\frac{r_1}{r}-1)(\frac{r_p}{r}-1)(\frac{r_a}{r}-1)}}\d r.\label{Brtau}
\eea
Let $\frac{r}{r_a}=\sin^2\psi$, $\frac{r_1}{r_a}=\sin^2\psi_1$ and $\frac{r_p}{r_a}=\sin^2\psi_p$, then replace back after the integration, we obtain the expression for the proper time,
\bea
\frac{-\sqrt{1-E^2}}{r_a}\tau&=&\frac{\sqrt{r_p(r_a-r_1)}}{r_a}\cE(\arcsin(\sqrt{\frac{r(r_1-r_a)}{r_1(r-r_a)}})|\frac{r_1(r_p-r_a)}{r_p(r_1-r_a)})\\
&+&\frac{r_1+r_a}{\sqrt{r_p(r_a-r_1)}}\cF(\arcsin(\sqrt{\frac{r(r_1-r_a)}{r_1(r-r_a)}})|\frac{r_1(r_p-r_a)}{r_p(r_1-r_a)}),\\
&-&\frac{r_1+r_a+r_p}{\sqrt{r_p(r_a-r_1)}}\Pi(\frac{r_1}{r_1-r_a};\arcsin(\sqrt{\frac{r(r_1-r_a)}{r_1(r-r_a)}})|\frac{r_1(r_p-r_a)}{r_p(r_1-r_a)})\\
&+&\sqrt{\frac{r(r_1-r)(r_p-r)}{r_a^2(r_a-r)}},
\eea
where $\cF(x|c)$ is the elliptic integral of the first kind, $\cE(x|c)$ is the elliptic integral of the second kind , and $\Pi(n;x|c)$ is the incomplete elliptic integral of the third kind, which are defined in Appendix \ref{appA}.

Similarly, let $\frac{r_+}{r_a}=\sin^2\psi_+$ and $\frac{r_-}{r_a}=\sin^2\psi_-$, and replace back after the integration, we obtain the solution of the azimuthal motion,
\bea
\phi&=&\frac{2r_a}{\sqrt{1-E^2}\sqrt{r_1(r_a-r_p)}}(\frac{2(\ell-aE)-r_a\ell}{(r_a-r_-)(r_a-r_+)}\cF(\arcsin(\sqrt{\frac{r(r_p-r_a)}{r_p(r-r_a)}})|\frac{r_p(r_a-r_1)}{r_1(r_a-r_p)})\nn\\
&-&\frac{2(\ell-aE)-r_-\ell}{(r_a-r_-)(r_+-r_-)}\Pi(\frac{r_p(r_a-r_-)}{r_-(r_a-r_p)};\arcsin(\sqrt{\frac{r(r_p-r_a)}{r_p(r-r_a)}})|\frac{r_p(r_a-r_1)}{r_1(r_a-r_p)})\nn\\
&-&\frac{2(\ell-aE)-r_+\ell}{(r_a-r_+)(r_+-r_-)}\Pi(\frac{r_p(r_a-r_+)}{r_+(r_a-r_p)};\arcsin(\sqrt{\frac{r(r_p-r_a)}{r_p(r-r_a)}})|\frac{r_p(r_a-r_1)}{r_1(r_a-r_p)})).
\eea
Noticing that there exists a symmetry by exchanging $r_1$ and $r_p$ in \eqref{Brtau} and the replacements below \eqref{Brtau}, one can easily obtain the another equivalent solution as 
\bea
\phi&=&\frac{2r_a}{\sqrt{1-E^2}\sqrt{r_p(r_a-r_1)}}(\frac{2(\ell-aE)-r_a\ell}{(r_a-r_-)(r_a-r_+)}\cF(\arcsin(\sqrt{\frac{r(r_1-r_a)}{r_1(r-r_a)}})|\frac{r_1(r_a-r_p)}{r_p(r_a-r_1)})\nn\\
&-&\frac{2(\ell-aE)-r_-\ell}{(r_a-r_-)(r_+-r_-)}\Pi(\frac{r_1(r_a-r_-)}{r_-(r_a-r_1)};\arcsin(\sqrt{\frac{r(r_1-r_a)}{r_1(r-r_a)}})|\frac{r_1(r_a-r_p)}{r_p(r_a-r_1)})\nn\\
&-&\frac{2(\ell-aE)-r_+\ell}{(r_a-r_+)(r_+-r_-)}\Pi(\frac{r_1(r_a-r_+)}{r_+(r_a-r_1)};\arcsin(\sqrt{\frac{r(r_1-r_a)}{r_1(r-r_a)}})|\frac{r_1(r_a-r_p)}{r_p(r_a-r_1)})).\label{BTorbit}
\eea
Note that this solution also applies to the orbits inside the horizon.

\begin{figure}
     \centering
     \includegraphics[width=0.99\textwidth]{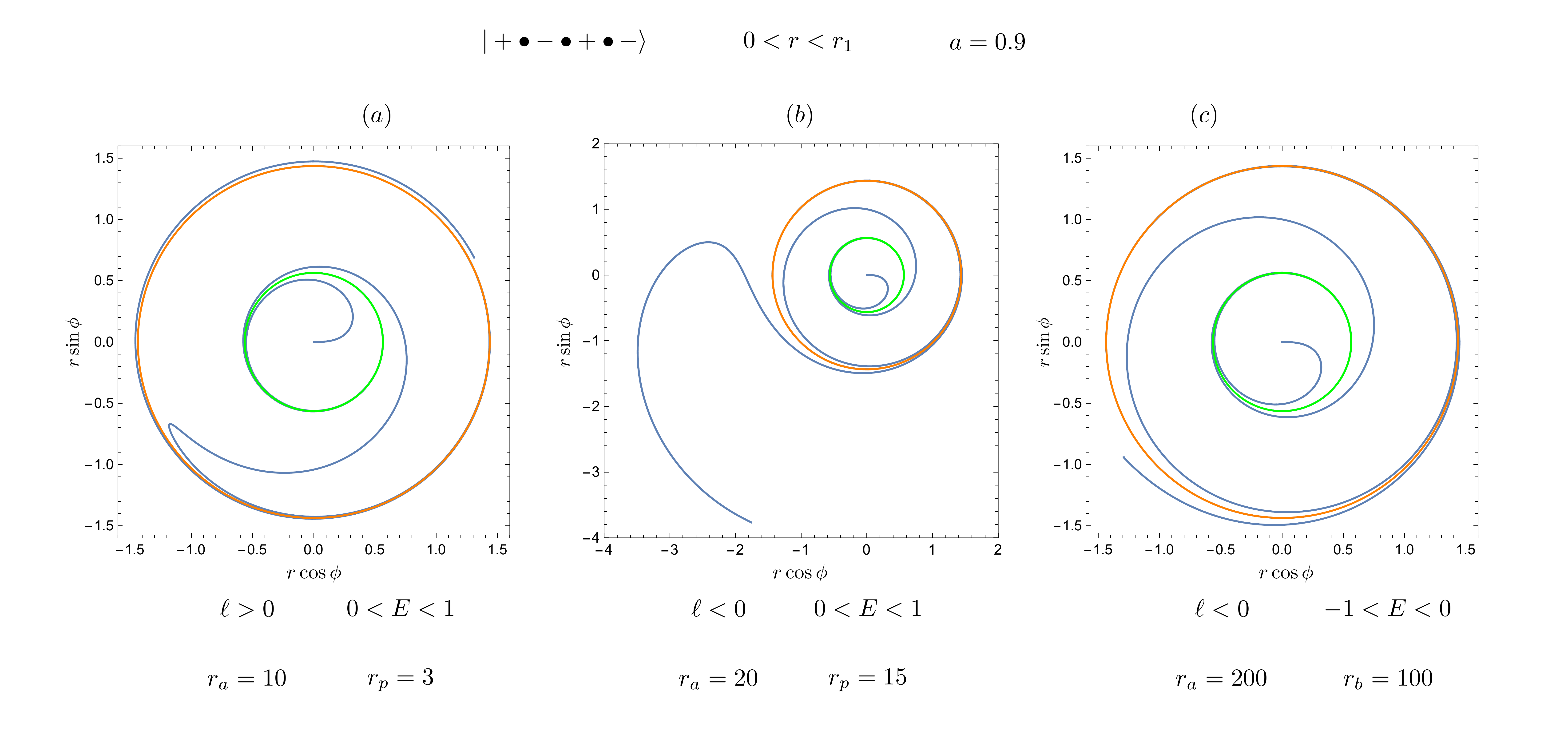}
     \caption{Geodesics related to the bound orbits with the root structure $\vert+\bullet-\bullet+\bullet-\hspace{2pt}\rangle$.}
     \label{bound}
 \end{figure}

In Figure \ref{bound}, we illustrate the behavior of the trapped orbits, which exhibits similarities with the trapped orbits related to circular orbits.

\subsection{Trapped Orbits Related to Deflecting Orbits}

The trapped orbits is confined in the first $+$ region of the root structure $\vert+\bullet-\bullet\hspace{2pt}+\rangle$, where we have $r_n<0<r_+<r<r_1<r_d<-r_n$. Setting $\frac{r}{-r_n}=\sin^2\psi$, $\frac{r_1}{-r_n}=\sin^2\psi_1$ and $\frac{r_d}{-r_n}=\sin^2\psi_d$, then 
\bea
-\sqrt{E^2-1}\d\tau=\frac{\sin2\psi}{\sqrt{(1+\csc^2\psi)(\csc^2\psi\sin^2\psi_1-1)(\csc^2\psi\sin^2\psi_1-1)}}\d\psi,
\eea
let $x=\csc\psi$, and replace back after the integration, we obtain the proper time expressed as
\bea
\frac{-\sqrt{E^2-1}}{(-r_n)}\tau&=&C_\cE\cE(\arcsin(\sqrt{\frac{(r-r_n)(r_n+r_d)}{(r+r_n)(r_d-r_n)}})|\frac{(r_1+r_n)(r_d-r_n)}{(r_d+r_n)(r_1-r_n)})\\
&+&C_\cF\cF(\arcsin(\sqrt{\frac{(r-r_n)(r_n+r_d)}{(r+r_n)(r_d-r_n)}})|\frac{(r_1+r_n)(r_d-r_n)}{(r_d+r_n)(r_1-r_n)}),\\
&+&C_\Pi\Pi(\frac{r_d-r_n}{r_n+r_d};\arcsin(\sqrt{\frac{(r-r_n)(r_n+r_d)}{(r+r_n)(r_d-r_n)}})|\frac{(r_1+r_n)(r_d-r_n)}{(r_d+r_n)(r_1-r_n)})\\
&+&\frac{\sqrt{(r-r_1)(r_n^2-r_1^2)(r-r_d)}}{2r_n(r+r_n)}.
\eea
where 
\bea
C_\cE=\frac{\sqrt{(r_n-r_1)(r_n+r_d)}}{-2r_n},\\
C_\cF=\frac{r_d}{\sqrt{(r_n-r_1)(r_n+r_d)}},\\
C_\Pi=\frac{-(r_1+r_d)}{\sqrt{(r_n-r_1)(r_n+r_d)}}.
\eea

Similarly we have the solution of the $\phi$ motion,

\bea
\phi&=&\frac{2r_n(2(\ell-aE)-r_n\ell)}{\sqrt{(E^2-1)(r_1-r_n)r_d}(r_n-r_-)(r_n-r_+)}\cF(\arcsin(\sqrt{\frac{r(r_1-r_n)}{r_1(r-r_n)}})|\frac{r_1(r_d-r_n)}{(r_1-r_n)r_d})\nn\\
&-&\frac{2r_n(2(\ell-aE)-r_-\ell)}{\sqrt{(E^2-1)(r_1-r_n)r_d}(r_n-r_-)(r_- -r_+)}\Pi(\frac{r_1(r_- -r_n)}{r_-(r_1-r_n)};\arcsin(\sqrt{\frac{r(r_1-r_n)}{r_1(r-r_n)}})|\frac{r_1(r_d-r_n)}{(r_1-r_n)r_d})\nn\\
&-&\frac{2r_n(2(\ell-aE)-r_+\ell)}{\sqrt{(E^2-1)(r_1-r_n)r_d}(r_+-r_-)(r_n -r_+)}\Pi(\frac{r_1(r_+ -r_n)}{r_+(r_1-r_n)};\arcsin(\sqrt{\frac{r(r_1-r_n)}{r_1(r-r_n)}})|\frac{r_1(r_d-r_n)}{(r_1-r_n)r_d}).\nn\\ \label{WTDef}
\eea
Note that the behavior of the trapped trajectory in this region is similar to the one depicted in Figure \ref{bound}. 

\subsection{Deflecting Orbits}
For the deflecting orbits in the second $+$ region from the second turning point to infinity, the redial velocity can be rewritten as
\bea
-\sqrt{E^2-1}\d\tau=\frac{1}{\sqrt{(1-\frac{r_1}{r})(1-\frac{r_n}{r})(1-\frac{r_d}{r})}}\d r.
\eea
After the integration, we obtain the proper time expressed as 
\bea
&&-\sqrt{E^2-1}\tau=\sqrt{\frac{r(r-r_n)(r-r_d)}{r-r_1}}-\cE(\arcsin(\sqrt{\frac{(r_n-r_1)(r-r_d)}{(r-r_1)(r_n-r_d)}})|\frac{r_1(r_d-r_n)}{r_d(r_1-r_n)})\nn\\
&+&\sqrt{\frac{1}{r_d(r_1-r_n)}}(r_1 (r_1 + r_n) + (r_1 - r_n) r_d)\cF(\arcsin(\sqrt{\frac{(r_n-r_1)(r-r_d)}{(r-r_1)(r_n-r_d)}})|\frac{r_1(r_d-r_n)}{r_d(r_1-r_n)})\nn\\
&-&\sqrt{\frac{1}{r_d(r_1-r_n)}}(r_1 - r_d) (r_1 + r_n + r_d)\Pi(\frac{r_d-r_a}{r_1-r_n};\arcsin(\sqrt{\frac{(r_n-r_1)(r-r_d)}{(r-r_1)(r_n-r_d)}})|\frac{r_1(r_d-r_n)}{r_d(r_1-r_n)})\nn .\\
\eea
For the $\phi$ motion, we have
\bea
\phi=\frac{2(\ell-aE)I_1^d-\ell I_2^d}{\sqrt{E^2-1}},
\eea
where
\bea
I^d_j=\int\frac{r^{j-1}}{\sqrt{\frac{(r-r_1)(r-r_n)(r-r_d)}{r}}(r-r_-)(r-r_+)}\d r .
\eea  
After the integration, we obtain the solution of the azimuthal motion, 
\bea
\phi&=&\frac{2(r_1-r_d)}{\sqrt{E^2-1}\sqrt{(r_1-r_n)r_d}}(\frac{2(\ell-aE)r_1-\ell r_1^2}{(r_1 - r_d) (r_1 - r_-) (r_1 - r_+)}\cF(\arcsin(\sqrt{\frac{(r_n-r_1)(r-r_d)}{(r-r_1)(r_n-r_d)}})|\frac{r_1(r_d-r_n)}{r_d(r_1-r_n)})\nn\\
&-&\frac{2(\ell-aE)r_--\ell r_-^2}{(r_- - r_d) (r_1 - r_-) (r_- - r_+)}\Pi(\frac{(r_d-r_n)(r_1-r_-)}{(r_1-r_n)(r_d-r_-)};\arcsin(\sqrt{\frac{(r_n-r_1)(r-r_d)}{(r-r_1)(r_n-r_d)}})|\frac{r_1(r_d-r_n)}{r_d(r_1-r_n)})\nn\\
&-&\frac{2(\ell-aE)r_+-\ell r_+^2}{(r_+ - r_d) (r_1 - r_+) (r_+ - r_-)}\Pi(\frac{(r_d-r_n)(r_1-r_+)}{(r_1-r_n)(r_d-r_+)};\arcsin(\sqrt{\frac{(r_n-r_1)(r-r_d)}{(r-r_1)(r_n-r_d)}})|\frac{r_1(r_d-r_n)}{r_d(r_1-r_n)})).\nn\\ \label{DeflectOut}
\eea

\begin{figure}
     \centering
     \includegraphics[width=0.99\textwidth]{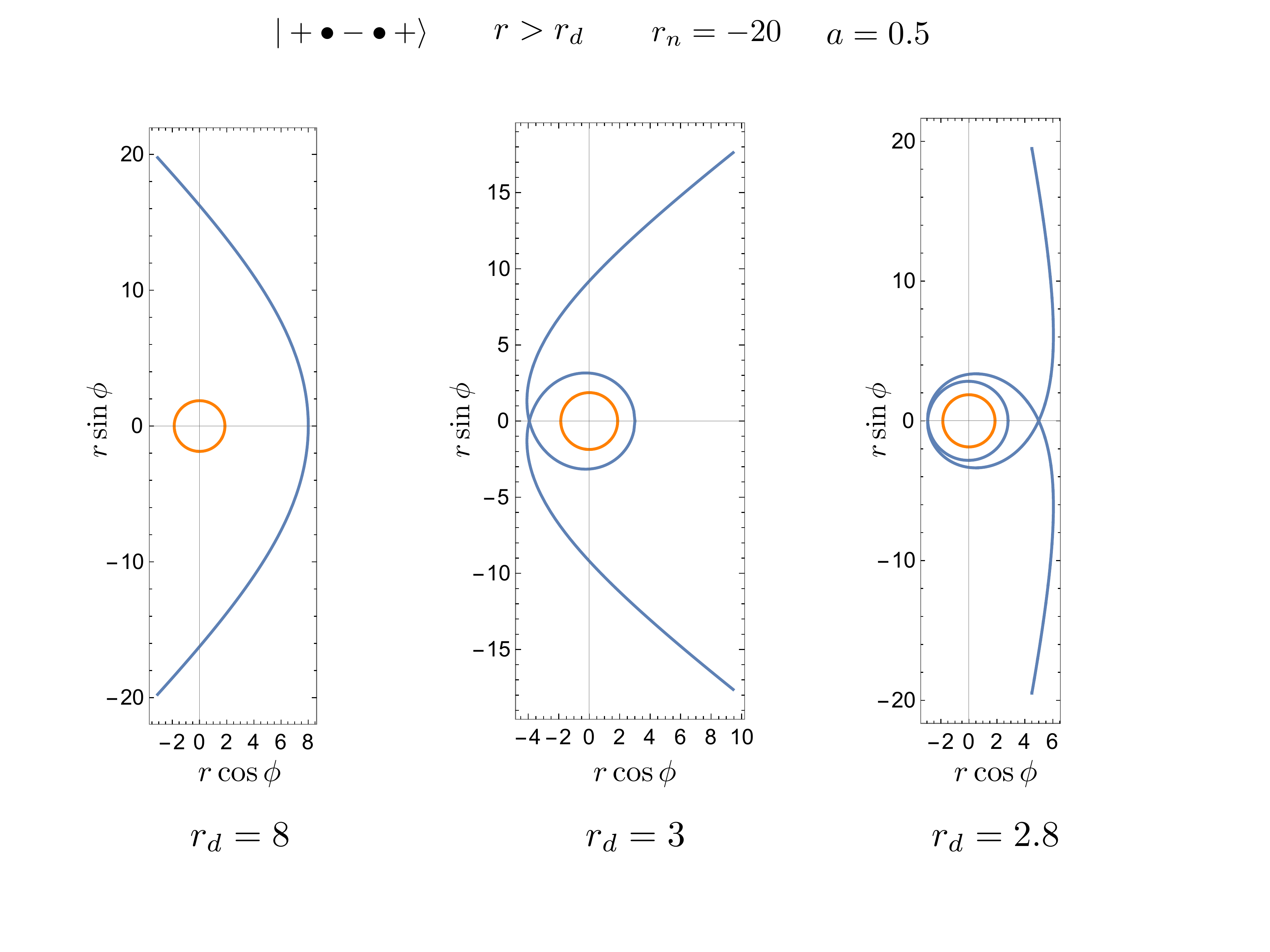}
     \caption{The deflecting orbits with the root structure $\vert+\bullet-\bullet\hspace{2pt}+\rangle$.}
     \label{Deflect}
 \end{figure}

In Figure~\ref{Deflect}, we depict the trajectory of deflecting orbits. An intriguing characteristic of these trajectories is that as the turning point approaches the turning point of the corresponding trapped orbits, the particle completes more revolutions around the black hole. When these two turning points merge into a double root, the orbit transforms into a whirling deflecting orbit, which asymptotically converges to the unstable circular orbit illustrated in Figure~\ref{UnReCirEp11}. Consequently, no outgoing trajectories are observed.

\section{Marginal Orbits with $E=1$ } \label{sec5}
For the radial potential of marginal orbits with $E=1$, there is one root going to infinity, then the radial potential is reduced to
\bea
R_1=2r^2-\ell^2r+2(a-\ell)^2.
\eea
The circular orbits locate at $r_*=r_c^{(1),(2)}$ in the root structure $\vert+\bullet\hspace{-4pt}\bullet+\rangle$, with the angular momentum
\bea
\ell_{M*}^{(1),(2)}=\pm2(1+\sqrt{1\mp a}),
\eea
the radial velocity is expressed as
\bea
U^r=\frac{\d r}{\d\tau}=\pm\frac{\sqrt{2r(r-r_*)^2}}{r^2}.
\eea
While for the root structure $\vert+\bullet-\bullet\hspace{2pt}+\rangle$ with the trapped and deflecting orbits, the turning point of the trapped orbit locates at $r_1=\frac{\ell^2}{2}-r_d$, where the angular momentum is 
\bea
\ell_{MD}^{(1),(2)}=\frac{-2a\pm\sqrt{2r_d\Delta(r_d)}}{r_d-2},
\eea
and the radial velocity is expressed as
\bea
U^r=\frac{\d r}{\d\tau}=\pm\frac{\sqrt{2r(r-r_1)(r-r_d)}}{r^2}.
\eea
When the second turning point locates at the ergosphere, i.e. $r_d=2$, we have 
\bea
r_1=\frac{(a^2-4)^2}{8a^2}, \quad\ell=\frac{2}{a}+\frac{a}{2}.
\eea
Note that when $r_d>2$, we have $\ell_{MD}^{(1)}>0$ and $\ell_{MD}^{(2)}<0$; when $r_+<r_d<2$, we have $\ell_{MD}^{(1),(2)}>0$, which means that inside the ergoregion, only the prograde marginal deflecting orbits exist. Note that when $E=-1$, the angular momentum is $-\ell_{MD}^{(1)}$.

\subsection{Unstable Circular Orbits}

For the whirling trapped orbits in the first $+$ region of the root structure $\vert+\bullet\hspace{-4pt}\bullet+\rangle$,
\bea
-\sqrt{2}\d\tau=\frac{\sqrt{r}}{\frac{r_*}{r}-1}\d r,
\eea
then after the integration we obtain
\bea
-\sqrt{2}\tau=2r_*^{3/2}\tanh^{-1}\sqrt{\frac{r}{r_*}}-\frac{2}{3}\sqrt{r}(r+3r_*).
\eea
For the whirling deflecting orbits in the second $+$ region of the root structure $\vert+\bullet\hspace{-4pt}\bullet+\rangle$,
\bea
-\sqrt{2}\d\tau=\frac{\sqrt{r}}{1-\frac{r_*}{r}}\d r,
\eea
then
\bea
-\sqrt{2}\tau=-2r_*^{3/2}\tanh^{-1}\sqrt{\frac{r_*}{r}}+\frac{2}{3}\sqrt{r}(r+3r_*).
\eea
The explicit expressions for $\phi$ motion are given as follows:
\begin{itemize}
    \item whirling trapped orbits in the region $r_+<r<r_*$,
        \bea
        \phi=C_*\tanh^{-1}\sqrt{\frac{r}{r_*}}+C_-\tanh^{-1}\sqrt{\frac{r_-}{r}}+C_+\tanh^{-1}\sqrt{\frac{r_+}{r}},
        \eea
    \item whirling deflecting orbits in the region $r>r_*$,
        \bea
        \phi=-C_*\tanh^{-1}\sqrt{\frac{r_*}{r}}-C_-\tanh^{-1}\sqrt{\frac{r_-}{r}}-C_+\tanh^{-1}\sqrt{\frac{r_+}{r}},
        \eea
    \item the orbits in the region $r_-<r<r_+$,
        \bea
        \phi=C_*\tanh^{-1}\sqrt{\frac{r}{r_*}}+C_-\tanh^{-1}\sqrt{\frac{r_-}{r}}+C_+\tanh^{-1}\sqrt{\frac{r}{r_+}},
        \eea
    \item the orbits in the region $0<r<r_-$,
        \bea
        \phi=C_*\tanh^{-1}\sqrt{\frac{r}{r_*}}+C_-\tanh^{-1}\sqrt{\frac{r}{r_-}}+C_+\tanh^{-1}\sqrt{\frac{r}{r_+}},
        \eea
\end{itemize}

where
\bea
C_*&=&\frac{\sqrt{r_*}(2(\ell-a)-r_*\ell)}{\sqrt{2}(r_--r_*)(r_+-r_*)},\\
C_-&=&\frac{\sqrt{r_-}(2(\ell-a)-r_-\ell)}{\sqrt{2}(r_+-r_-)(r_*-r_-)},\\
C_+&=&\frac{\sqrt{r_+}(2(\ell-a)-r_+\ell)}{\sqrt{2}(r_+-r_-)(r_+-r_*)}.
\eea

\subsection{Trapped and Deflecting Orbits}

When the angular momentum $|\ell|>\ell_{M*}^{(1)}$ or $\ell<\ell_{M*}^{(2)}$, the root structure is $\vert+\bullet-\bullet\hspace{2pt}+\rangle$. Note that similarly with the non-marginal deflecting orbits, when $E=1$ and $\ell_{M*}^{(1)}<\ell<\ell_+$ or $\ell<\ell_{M*}^{(2)}$, both the trapped orbits and the deflecting orbits exist; when $E=1$ and $\ell>\ell_+$, the trapped orbits are disallowed although the deflecting orbits exist; when $E=-1$ and $\ell<\ell_+$, the trapped orbits are allowed although the deflecting orbits do not exist.\\
For the trapped orbits in the first $+$ region of the root structure $\vert+\bullet-\bullet+\rangle$, $0<r<r_1<r_d$, let $\frac{r}{r_d}=x^2$, we obtain the proper time expressed as
\bea
\tau&=&-\frac{\sqrt{2}}{3}(\sqrt{r(r_1-r)(r_d-r)}-2\sqrt{r_1}(r_1+r_d)\cE(\arcsin\sqrt{\frac{r}{r_d}}|\frac{r_d}{r_1})\nn\\
&+&\sqrt{r_1}(2r_1+r_d)\cF(\arcsin\sqrt{\frac{r}{r_d}}|\frac{r_d}{r_1})).
\eea
The $\phi$ motion for trapped orbits is expressed as
\bea
\phi&=&-\sqrt{\frac{2}{r_d}}\ell\cF(\arcsin\sqrt{\frac{r}{r_1}}|\frac{r_1}{r_d})+\frac{\sqrt{2}(2a+\ell(r_--2))}{\sqrt{r_d}(r_--r_+)}\Pi(\frac{r_1}{r_-};\arcsin\sqrt{\frac{r}{r_1}}|\frac{r_1}{r_d})\nn\\
&-&\frac{\sqrt{2}(2a+\ell(r_+-2))}{\sqrt{r_d}(r_--r_+)}\Pi(\frac{r_1}{r_+};\arcsin\sqrt{\frac{r}{r_1}}|\frac{r_1}{r_d}).
\eea
For the deflecting orbits in the second $+$ region of the root structure $\vert+\bullet-\bullet+\rangle$, we introduce $\frac{r_1}{r}=\sin\psi$ and $\frac{r_1}{r_d}=\sin\psi_1$ to confine the variables in the deflecting region, then we obtain the proper time expressed as
\bea
-\tau&=&\frac{\sqrt{2}}{3r}\sqrt{r(r-r_1)(r-r_d)}(r+2(r_1+r_d))\nn\\
&+&\frac{2\sqrt{2}}{3}\sqrt{r_1}(r_1+r_d)\cE(\arcsin\sqrt{\frac{r_1}{r}}|\frac{r_d}{r_1})\nn\\
&-&\frac{\sqrt{2}}{3}\sqrt{r_1}(2r_1+r_d)\cF(\arcsin\sqrt{\frac{r_1}{r}}|\frac{r_d}{r_1}).\label{Examplept}
\eea
For the $\phi$ motion of marginal deflecting orbits, \eqref{phir} is expressed as
\bea
\frac{\d\phi}{\d r}=\frac{2(\ell-a)-\ell r}{\sqrt{2}\sqrt{\frac{(r-r_1)(r-r_d)}{r}}(r-r_-)(r-r_+)}.
\eea
The direct integration of the equation above contains imaginary terms. To obtain the real solution, we split this integral into two parts such that
\bea
\phi=\frac{2(\ell-a)I_1-\ell I_2}{\sqrt{2}},
\eea
while $I_2$ contains the imaginary terms and is not real, $I_1$ can be expressed simply as 
\bea
I_1=\frac{\sqrt{2}}{\sqrt{r_d}(r_--r_+)}(\Pi(\frac{r_1}{r_-};\arcsin\sqrt{\frac{r}{r_1}}|\frac{r_1}{r_d})-\Pi(\frac{r_1}{r_+};\arcsin\sqrt{\frac{r}{r_1}}|\frac{r_1}{r_d})).
\eea
However, it is still not real due to the inverse sin function. By noticing the structure of the proper time solution in \eqref{Examplept}, and comparing with the results in previous sections, we guess that the solution might contain functions such as 
\bea
\Pi(\frac{r_-}{r_1};\arcsin\sqrt{\frac{r_1}{r}}|\frac{r_d}{r_1}),\quad \text{or}\quad  \Pi(\frac{r_+}{r_1};\arcsin\sqrt{\frac{r_1}{r}}|\frac{r_d}{r_1}),\quad  \text{and}\quad  \cF(\arcsin\sqrt{\frac{r_1}{r}}|\frac{r_d}{r_1}).
\eea
By comparing the derivatives of these functions to find the coefficients, we obtain the solution of the azimuthal motion expressed as
\bea
\phi&=&\frac{\sqrt{2}}{\sqrt{r_1}(r_--r_+)}(2a+\ell(r_--2))\Pi(\frac{r_-}{r_1};\arcsin\sqrt{\frac{r_1}{r}}|\frac{r_d}{r_1})\nn\\
&-&\frac{\sqrt{2}}{\sqrt{r_1}(r_--r_+)}(2a+\ell(r_+-2))\Pi(\frac{r_+}{r_1};\arcsin\sqrt{\frac{r_1}{r}}|\frac{r_d}{r_1}).
\eea

\section{Conclusion}\label{sec6}

In this paper, we provide explicit analytical solutions for the equatorial Kerr geodesics related to circular orbits, bound orbits, deflecting orbits, and marginal geodesics. Specifically, we focus on the region $\ell^u\leq \ell\leq \ell^s$ in the phase space depicted in Figure 8 of ~\cite{Compere:2021bkk}.

We investigate the geodesic motion in relation to circular motion, present the analytical solutions, and demonstrate the performance of the trajectories. We identify the turning points for the $\phi$ motion of retrograde trapped orbits, which depend on the parameters $a$ and $r_*$. Moreover, we provide general results for all retrograde trapped orbits. Additionally, we determine the positions of stable circular orbits, which serve as a criterion for the admissibility of related trapped orbits. We also analyze the trajectories of motions with negative energy and find that, despite negative angular momentum, the trajectories remain prograde outside the horizon.

Subsequently, we examine the trapped orbits associated with bound and deflecting orbits, revealing that although the trapped orbits may exhibit different mathematical expressions, their trajectory behaviors are quite similar. We further observe that as the radial turning point of the deflecting orbit approach the turning point of the related trapped orbit, the particle completes more circles around the black hole. When these two turning points merge into a double root, the resulting orbits become whirling deflecting orbits that asymptotically approach unstable circular orbits. Finally, we provide explicit expressions for marginal orbits, noting that only prograde marginal deflecting orbits can traverse the ergoregion.

From a theoretical perspective, trapped orbits are expected to originate from white holes and then plunge into black holes. However, in practice, such trajectories may arise from particle collisions, where the resulting particles have the opportunity to occupy the corresponding trapped or deflecting regions in phase space. Thus, our findings have potential implications for investigating collisional Penrose processes and scattering problems. Furthermore, some of the techniques we employ to obtain solutions may prove useful for solving non-equatorial geodesics.

\section*{Acknowledgments}
We would like to thank Dr. Jie Jiang, Dr. Chen Lan and Dr. Andrew Mummery for helpful discussions.  Y. L. is financially supported by Natural Science Foundation of Shandong Province under Grants No.ZR2023QA133 and Yantai University under Grants No.WL22B218.
B.S. is supported by the National Natural Science Foundation of China under Grants No. 12375046 and Beijing University of Agriculture under Grants No.QJKC-2023032.

\appendix

\section{A brief introduction on elliptic functions}\label{appA}
The elliptic functions are introduced for solving integrals in the form of\cite{f3b6f360-404e-3ba9-86db-8759d8ce66a9,Whittaker_Watson_1996}
\bea
\int F(x,\sqrt{R(x)})\d x\label{ply}
\eea
where $F(x,\sqrt{R(x)})$ is a rational function of $x$ and $R(x)$, and $R(x)$ is a cubic or quartic polynomial
\bea
R(x)=Ax^4+Bx^3+Cx^2+Dx+E,
\eea
where $A, B, C, D$ and $E$ are constants. It has been shown that a general elliptic integral can be expressed by three elliptic integrals\cite{f3b6f360-404e-3ba9-86db-8759d8ce66a9,Whittaker_Watson_1996}, the Legendre elliptic integrals of the first, second and third kind, which are defined as
\bea
\cF(\varphi|m)&=&\int^{\sin\varphi}_0\frac{1}{\sqrt{(1-t^2)(1-mt^2)}}\d t\\
              &=&\int^\varphi_0\frac{1}{\sqrt{1-m \sin^2\theta}}\d\theta,\\
\cE(\varphi|m)&=&\int^{\sin\varphi}_0\frac{1-mt^2}{\sqrt{(1-t^2)(1-mt^2)}}\d t\\
              &=&\int^\varphi_0\sqrt{1-m \sin^2\theta}\d\theta, \\
\Pi(n;\varphi|m)&=&\int^{\sin\varphi}_0\frac{1}{(1-nt^2)\sqrt{(1-t^2)(1-mt^2)}}\d t\\
                &=&\int^\varphi_0\frac{1}{1-n \sin^2\theta}\frac{1}{\sqrt{1-m \sin^2\theta}}\d\theta
\eea

The inverse of the elliptic integral of the first kind gives the elliptic function, namely the Weierstrass elliptic function or Jacobian elliptic function, by rewriting the polynomial into Weierstrass form or Legendre form. The Weierstrass elliptic function has been recently used to solve for non-equatorial Kerr geodesic motion \cite{Cieslik:2023qdc}. And with Jacobian elliptic function, the bound orbits and orbits related to spherical orbits are solved \cite{Fujita:2009bp,vandeMeent:2019cam}. 

Now consider the integral 
\bea
\int \frac{1}{\sqrt{(x-a)(x-b)(x-c)(x-d)}}\d x,\label{eg2}
\eea
where $a>b>c>d$. When $x\geq a$ or $x\leq d$, by taking the replacement 
\bea
t^2=\frac{(x-a)(b-d)}{(x-b)(a-d)}, \quad m=\frac{(a-d)(b-c)}{(a-c)(b-d)},
\eea
the integral \label{eg1} can be rewritten in the form of the elliptic integral of the first kind 
\bea
\int\frac{2}{\sqrt{(a-c)(b-d)}}\frac{1}{\sqrt{(1-t^2)(1-mt^2)}}\d t.
\eea
For more replacements under different cases, we refer readers to \cite{f3b6f360-404e-3ba9-86db-8759d8ce66a9}.

\section{Consistency check with bound and circular cases}\label{appB}
When the eccentricity of the bound orbits goes to zero,  the separated roots $r_p$ and $r_a$ merge into a double root, and the bound motion turns into a stable circular orbital motion locates at $r_p=r_a\to r_*$, then the elliptic functions in solution \eqref{BTorbit} becomes
\bea
\cF&\to&-\arctan\sqrt{\frac{r_*(r-r_1)}{r(r_1-r_*)}},\\
\Pi_-&\to&\sqrt{\frac{r_-(r_*-r_1)}{r_*(r_1-r_-)}}\tanh^-1\sqrt{\frac{r_-(r_1-r)}{r(r_1-r_-)}},\\
\Pi_+&\to&\sqrt{\frac{r_+(r_*-r_1)}{r_*(r_1-r_+)}}\tanh^-1\sqrt{\frac{r_+(r_1-r)}{r(r_1-r_+)}},
\eea
where $\cF$, $\Pi_-$ and $\Pi_+$ are the elliptic functions in the first, second and third term of \eqref{BTorbit} respectively. Combined with the coefficients, the trapped orbital motion associated with stable circular motion \eqref{ScirTr} is then obtained.

When the separated roots $r_p$ and $r_1$ merge into a double root, the unstable circular orbits emerges, and the trapped motion turns into the whirling trapped orbits, then the quantities in the solution \eqref{BTorbit} changes as 
\bea
r_1=r_p\to r_*, \quad r_a\to r_1
\eea
the elliptic integrals can be re-expressed as 
\bea
\cF &\to& \tanh^{-1}\sqrt{\frac{r(r_*-r_1)}{r_*(r-r_1)}},\\
\Pi_-&\to&\frac{(r_*-r_1)r_- \tanh^{-1}\sqrt{\frac{r(r_*-r_1)}{r_*(r-r_1)}}+\sqrt{r_*r_-(r_1-r_*)(r_1-r_-)\tanh^{-1}\sqrt{\frac{r(r_--r_1)}{r_-(r-r_1)}}}}{r_1(r_*-r_-)},\\
\Pi_+&\to&\frac{(r_*-r_1)r_+ \tanh^{-1}\sqrt{\frac{r(r_*-r_1)}{r_*(r-r_1)}}+\sqrt{r_*r_+(r_1-r_*)(r_1-r_+)\tanh^{-1}\sqrt{\frac{r(r_+-r_1)}{r_+(r-r_1)}}}}{r_1(r_*-r_+)}
\eea
Combined with the coefficients, one can easily obtain the solution of the corresponding whirling trapped solution \eqref{WTphi} related to unstable circular case. Likewise, for the deflecting orbits, when $r_1=r_d$, the deflecting orbits turns into the whirling deflecting orbits, and the whirling trapped orbits  \eqref{WTDef} associated with the deflecting orbits, become the whirling trapped orbits associated with the unstable circular orbits \eqref{WTUnCir}.

\section{Consistency check with non-equatorial deflecting case}\label{appC}
The authors in \cite{Cieslik:2023qdc} solved for the non-equatorial geodesic motion in Kerr spacetime in terms Weierstrass functions. Here we illustrate that our results matches very well, by performing equatorial outgoing deflecting orbits as an example. We chose the same value of the parameters as the second picture selected in Figure \ref{Deflect}. The corresponding parameters of the results in \cite{Cieslik:2023qdc} are as follows.
\bea
 \varepsilon=1.066951,\quad \lambda_z=3.785653, \quad\kappa=10.576662,\\
 \theta_0=\pi/2, \quad\epsilon_r=1, \quad\xi_0=3, \quad\varphi_0=0,\quad\delta=1.
\eea

 \begin{figure}
     \centering
     \includegraphics[width=0.99\textwidth]{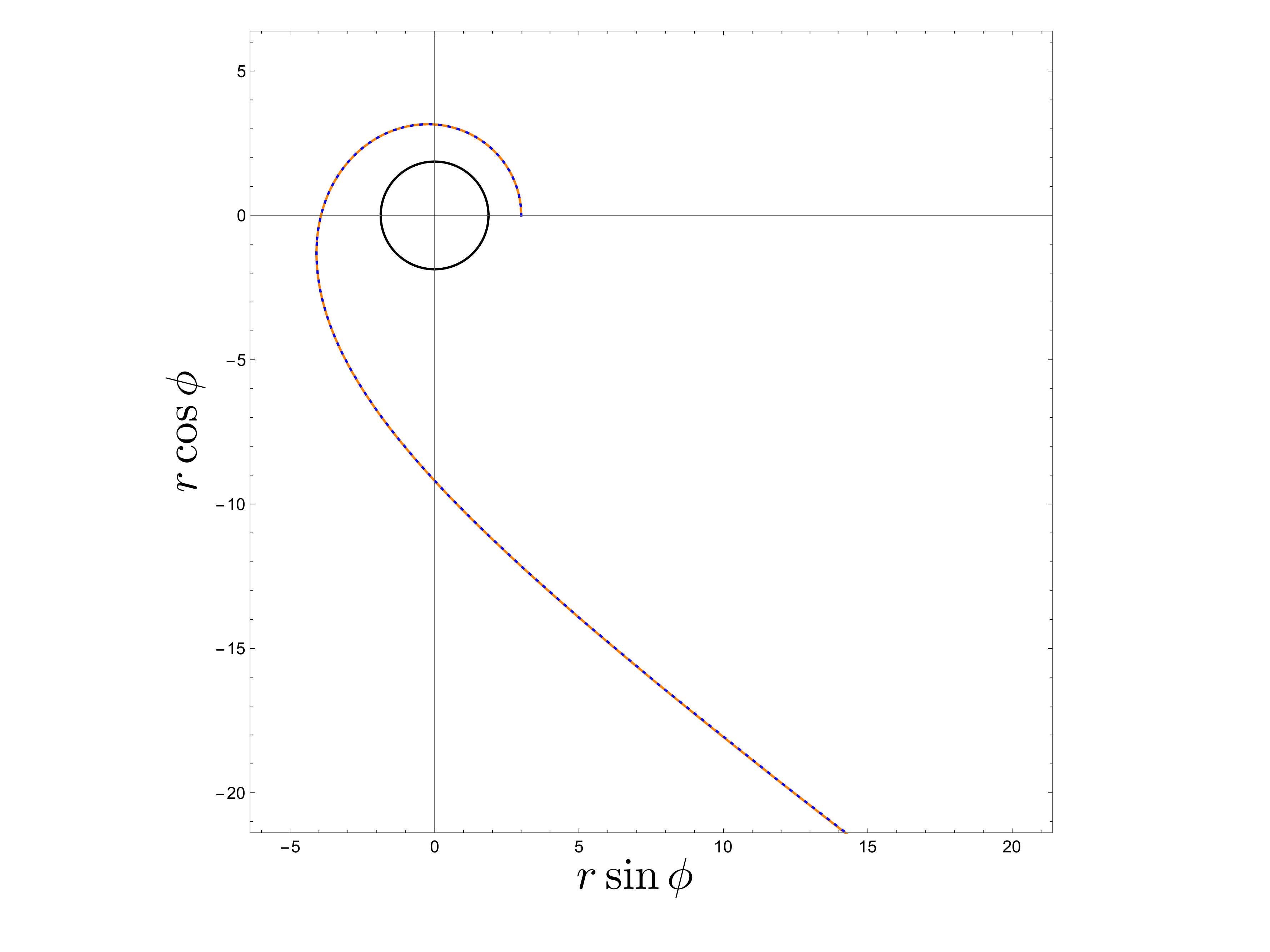}
     \caption{The comparison of outgoing deflecting orbits \eqref{DeflectOut}, which is in orange color, with the non-equatorial results in \cite{Cieslik:2023qdc}, which is the blue dashed curve. The black curve is the outer event horizon.}
 \end{figure}

\bibliography{refs}
\bibliographystyle{utphys}

\end{document}